\documentclass{article}
\usepackage[margin=1in]{geometry}
\usepackage{amsmath}
\usepackage{epsfig}
\numberwithin{equation}{section} % amsmath

\newcommand{\beq}{\begin{equation}}
\newcommand{\eeq}{\end{equation}}
\newcommand{\beqa}{\begin{eqnarray}}
\newcommand{\eeqa}{\end{eqnarray}}
\newcommand{\bdm}{\begin{displaymath}}
\newcommand{\edm}{\end{displaymath}}

\newcommand{\Eq}[1]{Eq.\ (\ref{#1})}
\newcommand{\Eqs}[2]{Eqs.\ (\ref{#1}) and (\ref{#2})}

\newcommand{\Rref}[1]{Ref.\ \cite{#1}}
\newcommand{\Rrefs}[2]{Refs.\ \cite{#1} and \cite{#2}}

\newcommand{\Fig}[1]{Fig.\ \ref{#1}}

\newcommand{\Section}[1]{Section\ \ref{#1}}
\newcommand{\Appendix}[1]{Appendix\ \ref{#1}}
\renewcommand{\Re}{\mbox{Re}\,}
\renewcommand{\Im}{\mbox{Im}\,}
\title{
Plasmon damping in a charged Bose-Einstein condensate model
}

\author{Jos\'e F. Nieves\footnote{nieves@ltp.uprrp.edu}\\
  Laboratory of Theoretical Physics, Department of Physics\\
  University of Puerto Rico, R\'{\i}o Piedras, Puerto Rico 00936
  \and\\[12pt]
  Sarira Sahu\footnote{sarira@nucleares.unam.mx}\\
  Instituto de Ciencias Nucleares\\
  Universidad Nacional Aut\'onoma de Mexico\\
  Circuito Exterior, C. U.\\
  A. Postal 70-543, 04510 Mexico DF, Mexico\\
}

\date{September 2025}

\begin{document}
\maketitle

\begin{abstract}
  In this work we consider the calculation of the imaginary part of 
  the dispersion relations of the propagating modes
  in a model of a charged scalar Bose-Einstein (BE) condensate, as well
  as the contribution to the imaginary part of the longitudinal component
  of the photon polarization tensor and the dielectric constant.
  In that model, two modes correspond to the transverse photon polarizations,
  while the other two modes are combinations of the longitudinal photon and
  the massive scalar field, which we denote as the $(\pm)$ modes.
  The dispersion relations of the transverse modes have the usual form for
  transverse photons in a plasma, and we do not consider here any further.
  In a previous work we determined the real part of the dispersion relations
  of the $(\pm)$ modes, as well as the real part of the longitudinal
  component of of polarization tensor and dielectric constant,
  which have some unique features. In the appropriate limit, those results
  reproduce the results obtained for the dielectric constant
  and dispersion relation in non-relativistic models of
  the BE condensation of charged scalars. Here we determine, in the same model,
  their imaginary part. The results can be useful in physical contexts
  involving the electrodynamics of a charged scalar BE condensate, and can
  serve as benchmark results for further exploration.
\end{abstract}

%end-preamble

\section{Introduction and motivation}
\label{sec:intro}

The properties of a particle that propagates in a medium,
such as its dispersion relation, depend crucially on the
composition and conditions of the background.
Models that extend the standard electroweak theory usally include
additional scalar particles, neutral and charged,
that interact with the standard model particles. Such new
scalar particles and interactions can have implications in 
nuclear, astrophysics, and cosmological plasma systems, 
which may contain a background of the
scalar particles\cite{baym:nstar,thorsson:kaon,schmitt:kaon,li:hion}.
Thus, for example, the intrinsic properties of neutrinos
and photons propagating in such a background medium, such as the dispersion
relations and/or electromagnetic properties, can be modified
in special ways that lead to significant and observable effects.
Under the appropriate conditions, the scalar background
may form a Bose-Einstein (BE) condensate. This is the case,
for example, in the models involved in the hypothesis that the
dark-matter is a self-interacting BE
condensate\cite{dm:boehmer,dm:garani,dm:craciun,dm:sikivie,dm:fan}.

Since the BE condensate is a very special state of matter, the properties
of the particles propagating in such a background, such as the dispersion
relations, are different from those in normal matter. The
propagation of photons in a medium that contains a BE condenssate
has been studied using various approaches and in various contexts,
for example the electrodynamics of fermions and
charged scalar bosons at low temperature\cite{dolgov},
heavy-ion collisions\cite{voskresensky}, exciton-polaritons\cite{kasprzak}, 
the BE condensation of photons\cite{kruchkov,mendonca} and the
Cerenkov effect\cite{slyu}, and the reduction of the photon
velocity\cite{hau,dutton,liu} in atomic gases.

Motivated by these developments, in a previous work\cite{ns:gbec}
we presented a model for the propagation of photons in a thermal
background that contains a BE condensate of a charged scalar.
The model is in turn based on a model we had considered earlier\cite{ns:fbec}
to study the propagation of fermions in a BE condensate.
The field theoretical method we used in \Rrefs{ns:gbec}{ns:fbec}
to treat the BE condensate, which we use here, has been discussed by various
authors\cite{weldon:phimu,filippi:phimu,schmitt:phimu,haber}.

In \Rref{ns:gbec} we determined the dispersion relations of the
propagating modes after the symmetry breaking associated with the BE
condensation in the model. In that model, two modes correspond to the
transverse photon polarizations, while the other two modes are combinations
of the longitudinal photon and the massive scalar field, which we denote
as the $(\pm)$ modes. While the dispersion relations of the transverse modes
have the usual form for transverse photons in a plasma, the dispersion
relations of the ($\pm$) modes reveal some unique features. In our
previous work cited above, we determined and analyzed the real part of the
dispersion relations of the $(\pm)$ modes, as well as the real part of
the contribution to the longitudinal component of polarization tensor
and the corresponding contribution to the dielectric constant.
In particular, in the appropriate limit, the dispersion
relation of the $(-)$ mode reduces to the expression
for the dispersion relation that has been obtained in
non-relativistic models of the BE condensation of charged
scalars\cite{2prb,3prb}.

The focus of the present work is the calculation of
the imaginary part of the dispersion relations,
as well as the corresponding contributions to the imaginary part
of the longitudinal component of polarization tensor
and the dielectric constant, in the same model.
These parameters are important because they
determine the energy loss and absoption effects when the system
is subject to external perturbations.
The main result is a set of explicit formulas for
those quantities, obtained by the one-loop
calculations of the self-energy of the propagating ($\pm$) modes
in the context of TFT. These results can be usfeul in the
context of the various applications that have been considered
in the references mentioned above, and in the works related
to dark-matter searches using materials with special optical
properties\cite{dmsearches}.

To illustrate some features of the results
we consider in some detail the long wavelength limit, including
specifically the high and low frequency limits, and obtained
simple explicit expressions for the damping rate as well as the
background contribution of the longitudinal polarization
tensor and dielectric constant in those cases. These results are obtained
in a general way, within the context of the model, but not tied to any
specific application, and as such can be useful and pave the way
for further development in various contexts and applications.
In the following sections we present the details of the calculations,
which we summarize as follows.

In \Section{sec:model} we summarize for reference the essential features
of the scalar BE condensation model we use, followed in \Section{sec:gmodel}
by a review the basic elements of the model with the coupling to the photon.
In \Section{sec:gmodel} we focus our atention to the relevant properties
regarding the $(\pm)$ modes, in particular the dispersion relations and related
quantities. In \Section{sec:propagators} we obtain the expressions
for the thermal propagators of the $(\pm)$ modes, and
in \Section{sec:1particle-approximation} we state precisely
the one-particle approximation that we used in the calculations of the
damping terms. In \Section{sec:impigen} we discuss the framework
we use for the calculation, including the couplings that
are required to perform the one-loop calculation of the damping terms, 
The details of the calculation itself are given in \Section{sec-damping}.
There we obtain the expression for the imaginary part of the longitudinal
component of polarization tensor and the dielectric constant,
as well as the decay rate of the $(+)$ mode into two
$(-)$ modes. Explicit formulas for those quanties are given
by considering limiting cases of the general expressions,
such as the long-wavelength limit, and in the high or low frequency limits.
Further discussion and examples are given in
\Section{sec:discussion}, in particular the correspondense
with the results obtained for the real part of the dielectric constant
and dispersion relation in non-relativistic models of the BE condensation
of charged scalars\cite{2prb}. Some details of the calculations are given in
the appendices. Our closing statements are given in \Section{sec:conclusions}.
In short, we believe that our work and results presented here
open the path for further exploration of the model, which can be relevant
for aplications related to the optical properties of this or similar systems.

\section{The BE condensate model}
\label{sec:model}

For reference we first summarize the essential features
of the scalar BE condensation model we use, and then
review the basic elements of the model with the coupling to the photon.

The scalar model for the $\phi$, discussed in detail in \Rref{ns:fbec},
is based on the Lagrangian
\beq
\label{Lphi}
L_\phi = (d^\mu\phi)^\ast(d_\mu\phi) - V_{\phi}\,,
\eeq
where
\beq
\label{dmu}
d_\mu \equiv \partial_\mu - i\mu u_\mu\,,
\eeq
and
\beq
\label{Vphi}
V_\phi = m^2\phi^\ast\phi + \lambda(\phi^\ast\phi)^2\,.
\eeq
The parameter $\mu$ is to be identified with the chemical potential
of the $\phi$, and $u^\mu$ is to be identified with
the velocity four-vector of the medium,
\beq
\label{u}
u^\mu = (1,\vec 0)\,,
\eeq
in the medium's own rest frame.

As discussed in \Rref{ns:fbec}
the motivation for considering this model is based on the following result:
the calculation of the effective potential $V_\text{eff}(T,\mu)$ for $\phi$
can be carried out in TFT using $\mu = 0$ in the partition
(and/or distribution) function, but using the
$\mu$-dependent Lagrangian density $L_\phi$ defined in \Eq{Lphi}.
Then neglecting the $T$-dependent terms (that is, at zero temperature),
$V_\text{eff}(0,\mu)$ is simply the $U$ potential,
given below in \Eq{U}.
In other words, calculations with the $\phi$ propagators, 
$\phi$ Lagrangian density and partition function for $\phi$,
\begin{equation}
Z = e^{-\beta H + \alpha Q}\,,
\end{equation}
give the same result as calculating them with the Lagrangian density
for $\phi^\prime$ obtained by using
$\partial_\mu \rightarrow D_\mu = \partial_\mu - i\mu u_\mu$ and using
\beq
Z = e^{-\beta H^\prime}\,.
\eeq
A simple proof is given in \Rref{ns:gbec}.

Expanding the $d$ term in \Eq{Lphi},
\beq
\label{Lmu}
L_\phi = (\partial^\mu\phi)^\ast(\partial_\mu\phi) +
i[\phi^\ast (v\cdot\partial\phi) - (v\cdot\partial\phi^\ast)\phi] - U(\phi)\,,
\eeq
where
\beq
\label{U}
U \equiv V_\phi - \mu^2\phi^\ast\phi =
-(\mu^2 - m^2)\phi^\ast\phi + \lambda(\phi^\ast\phi)^2\,,
\eeq
with
\beq
\label{vu}
v_\mu = \mu u_\mu\,.
\eeq
The case in which $\mu^2 < m^2$ corresponds
to a standard massive complex scalar with mass $m^2 - \mu^2$.
On the other hand, if $\mu^2 > m^2$, the minimum of the potential is
not at $\phi = 0$. In this situation $\phi$ develops a non-zero expectation
value and the $U(1)$ symmetry is broken, which is identified as
corresponding to the BE condensation phase.

We assume the situation is such that
\beq
\label{sbcondition}
\mu^2 > m^2\,.
\eeq
Then putting
\beq
\label{phi}
\phi = \frac{1}{\sqrt{2}}\left(\phi_0 + \phi_1 + i\phi_2\right)\,,
\eeq
where
\beq
\langle\phi\rangle \equiv \frac{1}{\sqrt{2}}\phi_0\,,
\eeq
is the minimum of
\beq
U_0 = -\frac{1}{2}(\mu^2 - m^2)\phi^2_0 + \frac{1}{4}\lambda\phi^4_0\,,
\eeq
gives
\beq
\label{phi0}
\phi^2_0 = \frac{\mu^2 - m^2}{\lambda}\,.
\eeq
The Lagrangian density for the scalar excitations $\phi_{1,2}$,
obtained by substituting \Eqs{phi}{phi0} in \Eq{Lphi}, becomes
\beq
L_\phi = \frac{1}{2}
\left[(\partial^\mu\phi_1)^2 + (\partial^\mu\phi_2)^2\right] +
\phi_2 v\cdot\partial\phi_1 - \phi_1 v\cdot\partial\phi_2 - U(\phi)\,,
\eeq
where
\beq
U(\phi) = -\frac{1}{2}(\mu^2 - m^2)[(\phi_0 + \phi_1)^2 + \phi^2_2]
+ \frac{1}{4}\lambda[(\phi_0 + \phi_1)^2 + \phi^2_2]^2\,.
\eeq
Using \Eq{phi0}, it follows that only $\phi_1$ appears in
the quadratic part,
\beq
U_2 = \frac{1}{2} m^2_1\phi^2_1\,,
\eeq
with
\beq
\label{m1}
m^2_1 = 2(\mu^2 - m^2)\,,
\eeq
while $\phi_2$ does not appear. However, $\phi_1$ and $\phi_2$ are mixed
by the $v^\mu$ term.

A convenient way to diagonalize the bilinear part of $L_\phi$ and
determine the modes that have definite dispersion relations,
is to use matrix notation and write
\beq
\hat\phi = \left(
\begin{array}{l}
  \phi_1 \\ \phi_2
\end{array}\right)\,.
\eeq  
In momentum space, the bilinear part of $L_\phi$ can then be written in the form
\beq
L^{(2)}_\phi(k) = \frac{1}{2}
\hat\phi^\ast(k)\Delta^{-1}_\phi(k)\hat\phi(k)\,,
\eeq
where
\beq
\label{Deltainv}
\Delta^{-1}_\phi(k) = \left(
\begin{array}{ll}
  k^2 - m^2_1 & 2iv\cdot k\\
  -2iv\cdot k & k^2
\end{array}
\right)\,.
\eeq
The classical field equations are
\beq
\label{becmodeleqmotion}
\Delta^{-1}_\phi(k)\hat\phi = 0\,,
\eeq
and the dispersion relations of the eigenmodes are obtained by solving
\beq
\label{dreq0}
(\omega^2 - \kappa^2)(\omega^2 - \kappa^2 - m^2_1) - 4\mu^2\omega^2 = 0\,.
\eeq
In \Eq{dreq0} we have used \Eq{vu} and defined the variables
\beqa
\label{omegakappadef}
\omega & = & u\cdot k\,,\nonumber\\
\kappa & = & \sqrt{\omega^2 - k^2}\,.
\eeqa
In the rest frame of the medium
\beq
\label{krestframe}
k^\mu = (\omega,\vec\kappa)\,,
\eeq
with $\kappa = |\vec\kappa|$. The dispersion relations
can be written in the form
\beq
\label{omegapmfinal}
\omega^2_{\pm}(\kappa) = \kappa^2 + \frac{1}{2} m^2_{\rho} \pm
\sqrt{\frac{1}{4} m^4_{\rho} + 4\mu^2\kappa^2}\,,
\eeq
where
\beq
\label{becmodelmasses}
m^2_{\rho} = m^2_1 + 4\mu^2 = 6\mu^2 - 2m^2\,.
\eeq
It follows that
\beqa
\label{omegapmzerokappa}
\omega_{+}(0)  & = & m_\rho\,,\nonumber\\
\omega_{-}(0)  & = & 0\,.
\eeqa
The combinations of $\phi_{1,2}$ that have the definite dispersion
relations $\omega_{\pm}$ can be obtained from \Eq{becmodeleqmotion},
but we will not need them here and therefore not discuss them further.
We mention that, while the mode corresponding to $\omega_{-}$
is the realization of the Goldstone mode associated
with the breaking of the global $U(1)$ symmetry,
its dispersion relation is not the usual $\omega = \kappa$,
but the one given by $\omega_{-}(\kappa)$ in \Eq{omegapmfinal}.

\section{Charged BE condensate model}
\label{sec:gmodel}

As discussed in \Rref{ns:gbec}, to study the propagation
of photons in the BE condensate background, we extended
the model by considering a $\phi$ field that is electrically
charged and interacts with electromagnetic field in the standard way.
The Lagrangian density for the model is then
\beq
L = -\frac{1}{4}F^2 + L_{\phi A}\,,
\eeq
where $F_{\mu\nu}$ is the electromagnetic field tensor,
\beq
\label{Lphigauged}
L_{\phi A} = (\hat D^\mu\phi)^\ast(\hat D_\mu\phi) -  V_\phi\,,
\eeq
with
\beq
\label{Dgauged}
\hat D_\mu = d_\mu + iqA_\mu = \partial_\mu - iv_\mu + iqA_\mu\,,
\eeq
while $V_\phi$ and $v^\mu$ are defined in \Eqs{Vphi}{vu}, respectively.
Notice that, while it would seem that the presence of $\mu$
in the Lagrangian can be eliminated by letting
\beq
A_\mu = A^\prime_\mu - \frac{1}{q} v_\mu\,,
\eeq
this would actually restore the presence of $\mu$ in the partition function,
which defeats the purpose of using the transformed field $\phi^\prime$.

For our purposes, it is more convenient to adopt the unitary gauge
and parametrize $\phi$ in the form
\beq
\phi = \frac{1}{\sqrt{2}}(\phi_0 + \rho)e^{i\theta/\phi_0}\,.
\eeq
The field $\theta$ does not appear in $V_\phi$, and
by a gauge transformation it disappears also from the kinetic term.
To indicate clearly that we are using the unitary gauge, we denote
by $V_\mu$ the transformed vector potential
$V_\mu = A_\mu + \frac{1}{q\phi_0}\partial_\mu\theta$.
With this notation, $L_{\phi A}$ is written in the form
\beq
\label{LAphi4}
L_{\phi A}  = \frac{1}{2}(\partial\rho)^2 +
\frac{1}{2}(v - qV)^\mu(v - qV)_\mu(\phi_0 + \rho)^2 - V_\phi\,.
\eeq
Proceeding as in the previous section, in particular assuming again that
\beq
\mu^2 > m^2\,,
\eeq
it follows that $\phi_0$ is given as in \Eq{phi0}.
Furthermore, the bilinear part of $L$, including the $F^2$ term, is given by
\beq
\label{Lbilinearx}
L^{(2)} = -\frac{1}{4}F^2 + \frac{1}{2}m^2_V V^2 +
\frac{1}{2}(\partial\rho)^2 - \frac{1}{2}m^2_1\rho^2 -
2m_V (v\cdot V)\rho\,,
\eeq
where
\beq
\label{mV}
m^2_V = q^2\phi^2_0 = \frac{q^2}{\lambda}(\mu^2 - m^2)\,,
\eeq
and $m_1$ is defined in \Eq{m1}.

The physical picture is that we end up with the two fields $V$ and $\rho$.
The field $\theta$ becomes the longitudinal component of $V$,
which is mixed with the scalar field $\rho$, by the term $\rho v\cdot V$
in the Lagrangian. As a consequence of this,
the propagating modes with definite dispersion relations are
superpositions of the longitudinal component of $V$ and the $\rho$.
Finding the dispersion relations, and the corresponding \emph{mixing}
combinations of the longitudinal component of $V$ and the $\rho$ was the
subject of \Rref{ns:gbec}. Our goal here is to determine the imaginary part,
or damping term, of the dispersion relations.

For this purpose, below we summarize the salient features of the results
we obtained for the dispersion relations, and more importantly for the
details of the present work, the corresponding mixing parameters
that determine the propagating modes in terms of the fields $\rho$
and the longitudinal component of $V$.

\subsection{Mixing equations}

In momentum space $L^{(2)}$ becomes
\beq
\label{Lbilineark}
L^{(2)}(k) = -V^{\ast\mu}[k^2\tilde g_{\mu\nu} - m^2_V g_{\mu\nu}]V^\nu +
\rho^\ast(k^2 - m^2_1)\rho -
2\mu m_V[(u\cdot V)^\ast\rho + c.c]\,,
\eeq
where
\beq
\tilde g_{\mu\nu} = g_{\mu\nu} - \frac{k_\mu k_\nu}{k^2}\,.
\eeq
The first two terms are the usual terms for a vector with squared
mass $m^2_V$ and a scalar with squared mass $m^2_1$,
while the last term mixes the scalar with the longitudinal
components of $V^\mu$. We decompose $V^\mu$ in the form
\beq
\label{VTLk}
V^\mu = V^\mu_T + V_L e^\mu_{3} + \frac{k^\mu}{\sqrt{k^2}}V_k\,,\qquad
V^\mu_T = \sum_{i= 1,2}V_i e^\mu_i
\eeq
where
\beqa
e^\mu_{1,2} & = & (0,\vec e_{1,2})\,,
\qquad(\vec e_{1,2}\cdot\vec\kappa = 0)\,,\nonumber\\
e^\mu_3 & = & \frac{\tilde u^\mu}{\sqrt{-\tilde u^2}}\,,
\eeqa
with
\beqa
\label{utilde2}
\tilde u_\mu =
\tilde g_{\mu\nu} u^\nu & = & u_\mu - \frac{(k\cdot u)k_\mu}{k^2}\,,\nonumber\\
\sqrt{-\tilde u^2} & = & \frac{\kappa}{\sqrt{k^2}}\,.
\eeqa
Substituting \Eq{VTLk} in \Eq{Lbilineark}, it follows that
the component $V_k$ does not have a kinetic energy term, and therefore
it is not a dynamical variable. Eliminating it
from the Lagrangian by using its Lagrange equation,
\beq
\label{Vk}
V_k = \frac{2\mu(k\cdot u)}{m_V\sqrt{k^2}}\rho\,,
\eeq
and substituting it in \Eq{Lbilineark},
\beqa
\label{L2withoutVk}
L^{(2)}(k) & = & -(k^2 - m^2_V)V^\ast_T\cdot V_T +
(k^2 - m^2_V)V^{\ast}_L V_L +
\left(k^2 - m^2_1 - \frac{4\mu^2(k\cdot u)^2}{k^2}\right)\rho^\ast\rho
\nonumber\\
&&\mbox{} + 2\mu m_V \sqrt{-\tilde u^2}\left[V^\ast_L\rho + c.c.\right]\,.
\eeqa
Thus, the transverse modes $V_T$ have a standard degenerate
dispersion relation for the two polarizations, and we will not
consider them further.

The remaining terms mix the longitudinal mode
$V_L$ ith the scalar $\rho$. Introducing the notation
\beq
\xi = \left(\begin{array}{c}\rho\\ V_L\end{array}\right)\,,
\eeq
they can be written in the form
\beq
\label{LrhoVL}
L^{(2)}_{\xi}(k) = \xi^\dagger\left(\Delta^{(\xi)}_{F0}\right)^{-1}\xi\,,
\eeq
where
\beq
\label{rhoVpropagator}
\left(\Delta^{(\xi)}_{F0}\right)^{-1} = \left(\begin{array}{cc}
  k^2 -  m^2_{\rho} - \frac{4\mu^2\kappa^2}{k^2} &
  \frac{2\mu m_V\kappa}{\sqrt{k^2}}
  \\[12pt]
  \frac{2\mu m_V\kappa}{\sqrt{k^2}} & k^2 - m^2_V
\end{array}\right)\,.
\eeq
In writing \Eq{rhoVpropagator}  we have used the
relation $(k\cdot u)^2 = k^2 + \kappa^2$ as well as \Eq{becmodelmasses}.
The field equations, which determine
the dispersion relations and mixing parameters are
\beq
\label{VLrhoeqs}
\left(\Delta^{(\xi)}_{F0}\right)^{-1}\xi = 0\,.
\eeq

\subsection{Dispersion relations}
\label{subsec:dispersionrelations}

From \Eq{VLrhoeqs}, the equaion for the dispersion relations is
\beq
\label{dreq}
(k^2 - m^2_V)(k^2 - m^2_\rho) - 4\mu^2\kappa^2 = 0\,.
\eeq
Writing the momentum of each propagating mode $(s = \pm)$ as
\beq
k^\mu = (\omega^{(s)}_{r},\vec\kappa)\,,
\eeq
the dispersion relations are given by
\beq
\label{omegapm}
\omega^{(s)}_{r}(\kappa) = \sqrt{\kappa^2 + \sigma_{s}(\kappa)}\,,
\eeq
where
\beq
\label{sigmapm}
\sigma_{s}(\kappa) = 
\frac{1}{2}(m^2_{\rho} + m^2_V) + \frac{s}{2}\sigma(\kappa)\,,
\eeq
with
\beq
\label{sigma}
\sigma(\kappa) = 
\left[
(m^2_{\rho} - m^2_V)^2 + 16\mu^2\kappa^2\right]^{\frac{1}{2}}\,.
\eeq
As mentioned in \Section{sec:intro}, throughout this work we are assuming that
\beq
\label{Vrhohier}
m_{\rho} > m_{V}\,.
\eeq
In this case, at zero momentum,
\beq
\sigma(0) = m^2_\rho - m^2_V\,,
\eeq
and therefore
\beqa
\label{omegapmkappa0}
\omega^{(+)}_{r}(0) & = & m_{\rho}\,,\nonumber\\
\omega^{(-)}_{r}(0) & = & m_{V}\,.
\eeqa
This is an indication that, at $\kappa = 0$, the $(+)$ mode is a pure $\rho$
while the $(-)$ mode is a pure $V_L$, as we will confirm below.

\subsection{Mixing - wavefunctions}

Up to a normalization factor, the corresponding eigenvectors can be
written in the form
\beqa
\label{upm}
u_{+} & = &
\left(\begin{array}{c}
  \Lambda \\[12pt] -m_V\gamma_{+}
\end{array}\right)\,,
\nonumber\\[12pt]
u_{-} & = &
\left(\begin{array}{c}
  m_V\gamma_{-} \\[12pt] \Lambda + \gamma^2_{-}
\end{array}\right)\,.
\eeqa
where
\beqa
\label{gammaLambda}
\gamma_{\pm} & = & \frac{2\mu\kappa}{\sqrt{\sigma_{\pm}}}\,,\nonumber\\
\Lambda & = & \frac{1}{2}(m^2_{\rho} - m^2_V) + \frac{1}{2}\sigma\,.
\eeqa
In the limit $\kappa = 0$, these solutions reduce to
\beqa
\label{spinorskappazero}
u_{+} & \propto &
\left(\begin{array}{c}
  1 \\ 0
\end{array}\right)\,,\\
u_{-} & \propto &
\left(\begin{array}{c}
  0 \\ 1
\end{array}\right)\,,
\eeqa
In other words, at $\kappa = 0$ the $(+)$ mode is pure $\rho$
while the $(-)$ mode is a pure $V_L$, in agreement with
\Eq{omegapmkappa0}.

\subsection{Polarization tensor and longitudinal dielectric constant}
\label{sec:poltensor}

The contribution to the longitudinal dielectric constant can be
obtained is as follows. The procedure is to start from \Eq{L2withoutVk} 
and integrate out the variable $\rho$, which is the tree-level analog
of calculating the one-loop self-energy,
leaving only the dynamical variables of $V$, and then identifying the
polarization tensor by comparing the result with
\beq
\label{L2Veff}
L_{eff} = -V^{\ast\mu}\left[(k^2 - \pi^{(T)})R_{\mu\nu} +
(k^2 - \pi^{(L)})Q_{\mu\nu}\right]V^\nu\,,
\eeq
where $R_{\mu\nu}$ and $Q_{\mu\nu}$ are defined by
\beqa
Q_{\mu\nu} & = & \frac{\tilde u_\mu \tilde u_\nu}{\tilde u^2}\,,\nonumber\\
R_{\mu\nu} & = & \tilde g_{\mu\nu} - Q_{\mu\nu}\,.
\eeqa
In terms of the variables $V_{T,L}$ we have defined in \Eq{VTLk},
\beq
\label{LeffpiTL}
L_{eff} = -(k^2 - \pi^{(T)})V^\ast_T\cdot V_T 
+ (k^2 - \pi^{(L)} V^\ast_L V_L\,.
\eeq
In general, when the thermal corrections are taken into account,
$\pi^{(T,L)}$ develop an imaginaty part, which we indicate by writing
\beq
\pi^{(T,L)} = \pi^{(T,L)}_r + i\pi^{(T,L)}_i\,.
\eeq
Solving \Eq{VLrhoeqs} for $\rho$ and substituting the solution back
in \Eq{L2withoutVk} reproduces \Eq{LeffpiTL} with
\beq
\label{piL}
\pi^{(L)}_r \equiv m^2_V +
\frac{4\mu^2 m^2_V \kappa^2}{k^2(k^2 - m^2_{\rho}) - 4\mu^2\kappa^2}\,.
\eeq
The solutions of the equation
\beq
\label{dreqpiL}
k^2 - \pi^{(L)}_r = 0\,,
\eeq
reproduce the dispersion relations given in 
\Section{subsec:dispersionrelations}, as it should be.

For our current purposes, one merit of \Eq{piL} is that it gives
the contribution to the longitudinal dielectric function.
Using\cite{np:pisubpi},
\beq
\label{epsilonell}
\epsilon_\ell = 1 - \frac{\pi^{(L)}}{k^2}\,,
\eeq
we then obtain,
\beq
\label{Reepsilonell}
\Re\epsilon_\ell = 1 - \frac{m^2_V}{k^2}\left[
1 + \frac{4\mu^2 \kappa^2}{k^2(k^2 - m^2_{\rho}) - 4\mu^2\kappa^2}
\right]\,.
\eeq
As we discuss in some detail in \Section{sec:discussion}, in the
large $m_\rho$ limit, stated precsily there, the dispersion
relation of the $(-)$ mode, given by $\omega^{(-)}_r$ in \Eq{omegapm}, as
well as the expression for $\Re\epsilon_\ell$ given above in
\Eq{Reepsilonell}, reproduce the results obtained in non-relativistic
models of BE condensation of charged scalars\cite{2prb}.

The imaginary part of the longitudinal polarization tensor, which we denote
by $\pi^{(L)}_i$, arises at one-loop, and it is the subject of the present work.
The corresponding contribution to the imaginary part of the
longitudinal dielectric constant is then given by
\beq
\label{Imepsilonell}
\Im\epsilon_\ell = - \frac{\pi^{(L)}_i}{k^2}\,.
\eeq
The details of their calculation are presented in the sections that follow.

\section{Propagators}
\label{sec:propagators}

\subsection{Zero temperature $(\pm)$ propagators}

The zero temperature propagators of the $s = (\pm)$ modes
can be written in the standard form
\beq
\label{propagatorszerotemp}
\Delta^{(s)}_{F0}(k) = \frac{1}{k^2 - \sigma_s(\kappa) + i\epsilon}\,,
\eeq
where the $\sigma_s(\kappa)$ is given in \Eq{sigmapm}.
For notational purposes in the discussions that follow,
it is convenient to generalize the expression above
and write the propagator for any momentum vector (e.g., $q$) in the form
\beq
\Delta^{(s)}_{F0}(q) = \frac{1}{q^2 - \hat\sigma_s(q) + i\epsilon}\,,
\eeq
where
\beq
\hat\sigma_s(q) = \sigma_s(Q)\,,
\eeq
and
\beq
Q = \sqrt{(q\cdot u)^2 - q^2}\,.
\eeq

\subsection{Finite temperature $(\pm)$ propagators}

The damping of each mode $(s = \pm)$ is obtained from the imaginary
part of the self-energy, which in turn is determined
from the $(\pi_{s})_{12}$ component of the thermal self-energy of each mode,
calculated, in our case, to 1-loop.
One simplifying feature of the $(\pm)$ propagators is the fact that
$\sigma_s$ is function of $Q$ but not of $q\cdot u$,
and therefore the usual derivation of the thermal propagators can be applied
here as well. The relevant components of the thermal propagator matrices,
which enter in the 1-loop calculation of $(\pi_{s})_{12}$, are
\beqa
\label{tftpropagators}
\Delta^{(s)}_{21}(q) & = & -2\pi i\delta(q^2 - \hat\sigma_{s}(q))
\left[\eta_B(q) + \theta(q)\right]\,,\nonumber\\
\Delta^{(s)}_{12}(q) & = & -2\pi i\delta(q^2 - \hat\sigma_{s}(q))
\left[\eta_B(q) + \theta(-q)\right]\,,
\eeqa
%
%or equivalently
%
%\beqa
%\Delta^{(s)}_{21}(k) & = & -2\pi i\delta(k^2 - \sigma_{s}(\kappa))
%\left[\eta_B(k) + \theta(k)\right]\,,\nonumber\\
%\Delta^{(s)}_{12}(k) & = & -2\pi i\delta(k^2 - \sigma_{s}(\kappa))
%\left[\eta_B(k) + \theta(-k)\right]\,,
%\eeqa
%
where
\beqa
\eta_B(q) & = & \theta(k)n_B(x_q) + \theta(-q)n_B(-x_q)\,,
\eeqa
with
\beqa
x_q & = & \beta q\cdot u\,.
\eeqa
Further,
\beq
\theta(q) = H(q\cdot u)\,,
\eeq
where $H(z)$ is the step function, and $n_B$ is the
boson distribution function
\beq
\label{nzboson}
n_B(x) = \frac{1}{e^x - 1}\,.
\eeq

\subsection{Zero temperature flavor propagators}

From \Eq{rhoVpropagator} it follows that the flavor propagator is given by
\beq
\label{propagatorflavor1}
\Delta^{(\xi)}_{F0}(k) = \frac{C^{(\xi)}}{D^{(\xi)}}\,,
\eeq
where
\beq
C^{(\xi)} = \left(\begin{array}{ll}
  k^2 - m^2_V  & -\frac{2\mu m_V\kappa}{\sqrt{k^2}}
  \\[12pt]
  -\frac{2\mu m_V\kappa}{\sqrt{k^2}} &
    k^2 - m^2_{\rho} - \frac{4\mu^2\kappa^2}{k^2}
\end{array}\right)\,,
\eeq
and
\beqa
D^{(\xi)} & = & (k^2 - m^2_V)\left(k^2 - m^2_{\rho}
- \frac{4\mu^2\kappa^2}{k^2}\right) -
\frac{4\mu^2 m^2_V\kappa^2}{k^2}\nonumber\\
& = & (k^2 - m^2_V)(k^2 - m^2_{\rho}) - 4\mu^2\kappa^2\,.
\eeqa
The equation $D^{(\xi)} = 0$ gives us the dispersion relations that
we have already discussed, and in fact we can write
\beq
D^{(\xi)} = d_{+} d_{-}\,,
\eeq
where
\beq
d_s = k^2 - \sigma_{s}\,,
\eeq
and $\sigma_{s}$ has been defined in \Eq{sigmapm}.

Using the relations
\beqa
k^2 - m^2_V & = & d_{-} + \sigma_{-} - m^2_V\,,\nonumber\\
k^2 - m^2_{\rho} - \frac{4\mu^2\kappa^2}{k^2} & = &
d_{+} + \sigma_{+} - m^2_{\rho} - \frac{4\mu^2\kappa^2}{k^2}\,.
\eeqa
and
\beq
\frac{1}{d_{+}d_{-}} = \frac{1}{\sigma_{+} - \sigma_{-}}\left(
\frac{1}{d_{+}} - \frac{1}{d_{-}}\right) =
\frac{1}{\sigma}\left(
\frac{1}{d_{+}} - \frac{1}{d_{-}}\right)\,,
\eeq
the propagator in \Eq{propagatorflavor1} can be written in the form
\beq
\label{flavorpropagatorsfull}
\Delta^{(\xi)}_{F0}(k) = \sum_s \Delta^{(s)}_{F0}(k) C_s\,,
\eeq
where $\Delta^{(s)}_{F0}$ is given in \Eq{propagatorszerotemp}, and
\beqa
C_{+} & = & \frac{1}{\sigma}
\left(\begin{array}{ll}
  \sigma + \sigma_{-} - m^2_V  & -\frac{2\mu m_V\kappa}{\sqrt{k^2}}
  \\[12pt]
  -\frac{2\mu m_V\kappa}{\sqrt{k^2}} &
    \sigma_{+} - m^2_{\rho} - \frac{4\mu^2\kappa^2}{k^2}
\end{array}\right)\,\nonumber\\
C_{-} & = & \frac{1}{\sigma}
\left(\begin{array}{ll}
  -\left(\sigma_{-} - m^2_V\right)  & \frac{2\mu m_V\kappa}{\sqrt{k^2}}
  \\[12pt]
  \frac{2\mu m_V\kappa}{\sqrt{k^2}} &
    \sigma - \left(\sigma_{+} - m^2_{\rho} - \frac{4\mu^2\kappa^2}{k^2}\right)
\end{array}\right)\,.
\eeqa
Although we are not indicating it explicitly here, it should be
kept in mind that the the matrices $C_{\pm}$ are functions
of $\omega$ and $\kappa$.

\subsection{Finite temperature flavor propagators}

\Eq{flavorpropagatorsfull} is the formula for the full flavor
propagators at zero temperature. Their finite temperature
counterparts are given by the same expression but replacing
the zero temperature $\Delta^{(s)}_{F0}$ by their finite-temperature
counterparts, given in \Eq{tftpropagators}. Explicitly,
\beq
\label{tftflavorpropagators}
\Delta^{(\xi)}_{ab}(k) = \sum_s \Delta^{(s)}_{ab}(k) C_s
\eeq
where $a$ and $b$ stand for the thermal indices 1,2.

For sufficiently small $\kappa$, the $C_{\pm}$ matrices become
diagonal, 
$C_{+} \rightarrow \text{diag}(1,0), C_{-} \rightarrow \text{diag}(0,1)$,
and therefore
\beq
\Delta^{(\xi)}_{F0}(k) \rightarrow
\left(\begin{array}{cc}
  \Delta^{(+)}_{F0}(k) & 0\\
  0 & \Delta^{(-)}_{F0}(k)
\end{array}
\right)\,,
\eeq
and correspondingly for the thermal propagator matrix,
\beq
\label{thermalpropagatorsnomixing}
\Delta^{(\xi)}_{ab}(k) \rightarrow
\left(\begin{array}{cc}
  \Delta^{(+)}_{ab}(k) & 0\\
  0 & \Delta^{(-)}_{ab}(k)
\end{array}
\right)\,.
\eeq
That is, the flavor propagator matrix becomes diagonal
with $\rho$ identified with the $(+)$ mode and $V_L$ with the $(-)$ mode,
as we have already anticipated.

\subsection{One-particle approximation}
\label{sec:1particle-approximation}

The one-particle approximation consists in making the replacements
\beq
C_{s}(\omega,\kappa) \rightarrow C_{s}(\omega^{(s)}_{r},\vec\kappa)\,,
\eeq
in the expressions for the propagators in
\Eqs{flavorpropagatorsfull}{tftflavorpropagators}.
That is, aproximate the propagator by evaluating the numerators
at the dispersion relation. Thus, for example,
\beq
C_{+} \rightarrow C_{+}(\omega^{(+)}_{r},\kappa) = \frac{1}{\sigma}
\left(\begin{array}{ll}
  \sigma + \sigma_{-} - m^2_V  & -m_V\gamma_{+}
  \\[12pt]
  -m_V\gamma_{+} &
    \sigma_{+} - m^2_{\rho} - \gamma^2_{+}
\end{array}\right)\,,
\eeq
and similarly for $C_{-}(\omega^{(-)}_{r},\kappa)$, where $\gamma_{\pm}$
are given in \Eq{gammaLambda}. It is straightforward to confirm that
\beqa
\label{Cuu}
C_{+}(\omega^{(+)}_{r},\kappa) & = & \frac{1}{\Lambda\sigma}u_{+} u^\dagger_{+}
\,,\nonumber\\
C_{-}(\omega^{(-)}_{r},\kappa) & = & \frac{1}{(\Lambda + \gamma^2_{-})\sigma}
  u_{-} u^\dagger_{-}\,,
\eeqa
where $u_{\pm}$ are the (flavor) spinors given in \Eq{upm}.
In other words, in the one-particle approximation, the zero-temperature
flavor propagators are given by
\beq
\left(\Delta^{(\xi)}_{F0}(k)\right)_{1p} =
\frac{U_{+}U^\dagger_{+}}{k^2 - \sigma_{+} + i\epsilon}
+ \frac{U_{-}U^\dagger_{-}}{k^2 - \sigma_{-} + i\epsilon}\,,
\eeq
where
\beqa
U_{+} & = & \sqrt{\frac{1}{\Lambda\sigma}} u_{+}\,,\nonumber\\
U_{-} & = & \sqrt{\frac{1}{(\Lambda + \gamma^2_{-})\sigma}}
u_{-}\,.
\eeqa
The $U_{\pm}$ factors defined above are analogous in the present context
to the normalization factors for plasmons propagating in a plasma
or fermions propagating in a thermal background\cite{zaidi,weldon:fermions}.
They in turn determine the properly normalized
transformation matrix between the $\rho,V_L$ fields
and the fields of the eigenmodes. If we denote the latter by $\phi_{\pm}$, then
\beq
\label{fieldtransformation}
\left(\begin{array}{c}
  \rho\\ V_L
\end{array}\right) = U_{+}\phi_{+} + U_{-}\phi_{-}\,.
\eeq

The thermal propagators given in \Eq{tftflavorpropagators}
can be expressed as
\beq
\label{tftflavorpropagators-1part-ab}
\Delta^{(\xi)}_{ab}(k) = \sum_s \Delta^{(s)}_{ab}(k)U_s U^\dagger_s\,,
\eeq
in particular,
\beq
\label{tftflavorpropagators-1part}
\Delta^{(\xi)}_{12}(k) = \sum_s \Delta^{(s)}_{12}(k) U_s U^\dagger_s\,,
\eeq
and similarly for the (21) components, where the
$\Delta^{(\pm)}_{12}(k)$ and $\Delta^{(\pm)}_{21}(k)$ propagators are given
explicitly in \Eq{tftpropagators}.
The spinors $U_\pm$ that we have identified
above allow us to calculate the rate of processes such as
\beq
\label{decay+--}
\phi_{+} \rightarrow \phi_{-} + \phi_{-}\,,
\eeq
in a straightforward way once we know the vertices to use.
In turn, the thermal propagators allow us to calculate the thermal
corrections to such processes. The proper
vertices for such purposes are determined in \Section{subsec-vertices}.

\section{Imaginary part of the self-energy - Formulation of the problem}
\label{sec:impigen}

\subsection{General formulas for $\pi_i$}

Denoting by $\pi^{(\pm)}_r$ the real part of the self-energy of each
propagating mode, then from \Eq{propagatorszerotemp} they are simply
\beq
\label{pirpm}
\pi^{(\pm)}_r = \sigma_{\pm}\,.
\eeq
In general, including the loop corrections, the self-energy will develop
an imaginary part due to the interactions with the background.
Thus we write the total self-energy in the form
\beq
\pi^{(\pm)}_{eff}(k) = \pi^{(\pm)}_r(k) + i\pi^{(\pm)}_i(k)\,.
\eeq
In the case of one scalar field, a convenient formula to determine $\pi_i$ is
\beq
\label{piip12}
\pi_i(k) = \frac{i\pi_{12}(k)}{2n_B(x)}\,,
\eeq
where $\pi_{12}(k)$ is the 12 component of the thermal self-energy matrix.
The dispersion relation is the solution of
\beq
k^2 - (\pi_r + i\pi_i) = 0\,.
\eeq
Writing
\beq
\omega = \omega_r + i\omega_i\,,
\eeq
to the lowest order
\beq
\label{defGamma}
\omega_i(\kappa) = \frac{\pi_i(\omega_r,\kappa)}{2\omega_r N} \equiv
-\frac{\Gamma(\kappa)}{2}\,,
\eeq
where $\omega_r$ is the solution of
\beq
\omega^2_r - \pi(\omega_r,\kappa) = 0\,,
\eeq
and
\beq
\label{Nnormfactor}
N = 1 - \frac{\pi^\prime_r}{2\omega_r}\,,
\eeq
with
\beq
\pi^\prime_r \equiv \left.
\frac{\partial\pi_r}{\partial\omega}\right|_{\omega = \omega_r}\,.
\eeq
The quantity $\Gamma$ defined in \Eq{defGamma}
has the interpretation of the total decay rate;
the evolution amplitude of the propagating mode has the factor
\beq
\label{defDamping}
e^{-i\omega t} = e^{-i\omega_r t} e^{-\frac{\Gamma}{2}t}\,.
\eeq
and therefore the flux is damped by $e^{-\Gamma t}$.

In the spirit of the one-particle approximation,
similar formulas apply also to our case, for each propagating
$(s = \pm)$ mode. At this point it is useful to recall the fact
that $\sigma_{\pm}$, and whence $\pi^{(\pm)}_r$ by \Eq{pirpm}, are independent
of $\omega$ [\Eq{sigmapm}], and therefore the normalization factors
analogous to $N$ above are 1. Thus we write their dispersion relation
in the form
\beq
\omega_s = \omega^{(s)}_{r} - i\frac{\Gamma_s}{2}\,,
\eeq
with
\beq
\label{defGammamode}
\Gamma_s \equiv
-\frac{\pi^{(s)}_i(\omega^{(s)}_{r},\kappa)}{\omega^{(s)}_{r}}\,.
\eeq

The strategy now is to consider the cubic interaction terms
in the Lagrangian and write them in terms of the $\phi_{\pm}$ fields
by means of \Eq{fieldtransformation}. Then it becomes straighforward
to calculate the rate for processes such as
$\phi_{+} \rightarrow \phi_{-} + \phi_{-}$, and related ones,
as well as the thermal corrections from the one-loop diagrams for
the imaginary part of the self-energy of each mode.

\subsection{Vertices}
\label{subsec-vertices}

The relevant interaction terms are obtained from Eq. (3.8) of \Rref{ns:gbec},
\beq
\label{LAphi3}
L_{\phi A} = \frac{1}{2}(\partial\rho)^2 + \frac{1}{2}(q\phi_0)^2 V^2 +
\frac{1}{2}q^2 V^2\rho^2 + q^2\phi_0\rho V^2 -
2q\mu (u\cdot V)(\phi_0 + \rho)^2 - U_\phi\,.
\eeq
To write down the $V_L-\rho$ interaction terms we must eliminate the
component $V_k$ in the cubic and quartic terms, as we did it
for the bilinear terms, using \Eq{Vk}.
Since that formula depends on $k$, it will produce vertices
that are momentum-dependent. We consider the relevant
interactions terms that follow from \Eq{LAphi3} one by one.
The terms that must be looked at are
\beq
\label{Lint}
L^\prime = \frac{1}{2}q^2 V^2\rho^2 + q^2\phi_0\rho V^2 -
2q\mu(u\cdot V) \rho^2\,,
\eeq
where
\beqa
V^2 & = & V_T\cdot V_T - V^2_L - V^2_k \rightarrow
V_T\cdot V_T - V^2_L - \frac{4\mu^2(k\cdot u)^2}{m^2_V k^2}\rho^2
\,,\nonumber\\
u\cdot V & = & -\sqrt{-\tilde u^2} V_L + \frac{(k\cdot u)V_k}{\sqrt{k^2}}
\rightarrow -\sqrt{-\tilde u^2} V_L + \frac{2\mu(k\cdot u)^2}{m_V k^2}\rho\,.
\eeqa
Therefore, the $V_L-\rho$ interaction terms contained in \Eq{Lint}
are the following quartic and cubic terms,
\beqa
\label{L34}
L^{(4)} & = & -\frac{1}{2}q^2 V^2_L\rho^2\,,\nonumber\\
L^{(3)} & = & -q m_V\rho V^2_L + 2q\mu\sqrt{-\tilde u^2}V_L \rho^2\,,\nonumber\\
& = & -q m_V\rho V^2_L + q\gamma(k)V_L \rho^2\,,
\eeqa
where we have used
\beq
m_V = q\phi_0\,,
\eeq
and defined
\beq
\label{vrho2gamma}
\gamma(k) = \frac{2\mu\kappa}{\sqrt{k^2}}\,.
\eeq
\begin{figure}
\begin{center}
\epsfig{file=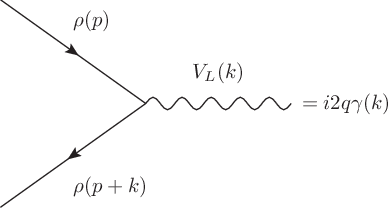,bbllx=235,bblly=348,bburx=423,bbury=448}
\end{center}
\caption[] {
  Feynman rule for the $V_L\rho^2$ vertex that appears in \Eq{L34},
  and $\gamma(k)$ defined in \Eq{vrho2gamma}.
  \label{fig1}
}
\end{figure}

\subsection{$L^{(3)}$ in terms of the eigenmode fields $\phi_{\pm}$}

For definiteness, we consider the calculation of $\pi^{(+)}_i$, that is,
the imaginary part of the self-energy of the $(+)$ mode, from which
we determine the corresponding damping as already explained.
As we know, the imaginary part of the self-energy is related to the
rates of various processes, and our focus on $\pi^{(+)}_i$ is
based on choosing the common case  in which the dominant processes
that contribute are $\phi_{+} \leftrightarrow \phi_{-} + \phi_{-}$.
The quartic term $V^2_L\rho^2$ can contribute to the diagonal ($11$ and $22$)
elements of the self-energy but not to the $12$ (or $21$) element, therefore
we do not consider them further. In short, the problem is to start from
\beq
L^{(3)} = -q m_V\rho V^2_L + q\gamma(k)V_L \rho^2\,,
\eeq
and write it in terms of $\phi_{\pm}$ using
\beq
\left(\begin{array}{c}
  \rho\\ V_L
\end{array}\right) = \sum_{i = \pm} U_i\phi_i\,.
\eeq
Thus writing
\beqa
\rho & = & \sum_i U_{1i} \phi_i\,,\nonumber\\
V_L & = & \sum_i U_{2i} \phi_i\,,
\eeqa
we have
\beq
L^{(3)} = -q m_V\sum_{ijk} U_{1i} U_{2j} U_{2k}\,
\phi_i \phi_j \phi_k +
q\gamma(k)\sum_{ijk} U_{2i} U_{1j} U_{1k}\,
\phi_i \phi_j \phi_k\,.
\eeq
Now we want to single out the term of the form $\phi_{+}\phi^2_{-}$\,.
This is,
\beq
\label{L3prime}
L^{\prime(3)} = A\phi_{+} \phi^2_{-}\,,
\eeq
where
\beqa
\label{A}
A & = & -q m_V (U_{1+} U_{2-} U_{2-} + U_{1-} U_{2+} U_{2-} +
U_{1-} U_{2-} U_{2+})\nonumber\\
&&\mbox{} +
q\gamma(k) (U_{2+} U_{1-} U_{1-} + U_{2-} U_{1+} U_{1-} +
U_{2-} U_{1-} U_{1+})\,.
\eeqa
To finish this step we have to use the explicit formulas for the
components $U_{ai}$ to determine the coupling $A$. The $U_{ai}$
are given explicitly by
\beqa
\label{Uai}
U_{1+} & = & \sqrt{\frac{1}{\Lambda\sigma}}\,\Lambda = 
\sqrt{\frac{\Lambda}{\sigma}}\,,\nonumber\\
U_{2+} & = & \sqrt{\frac{1}{\Lambda\sigma}}(-m_V\gamma_{+})\,,\nonumber\\
U_{1-} & = & \sqrt{\frac{1}{(\Lambda + \gamma^2_{-})\sigma}}(m_V\gamma_{-})
\,,\nonumber\\
U_{2-} & = & \sqrt{\frac{1}{(\Lambda + \gamma^2_{-})\sigma}}
(\Lambda + \gamma^2_{-}) = \sqrt{\frac{\Lambda + \gamma^2_{-}}{\sigma}}\,.
\eeqa
\begin{figure}
\begin{center}
\epsfig{file=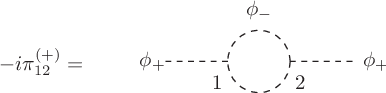,bbllx=196,bblly=383,bburx=382,bbury=428}
\end{center}
\caption[] {
  Diagrams for the 12 element of the self-energy of the $(+)$ mode
  due to the interaction term given in \Eq{L3prime}.
  \label{fig2}
}
\end{figure}

Using the vertex given in \Eq{L3prime}, we can calculate
$(\pi^{(+)}_{12})$ by the diagram given in \Fig{fig2},
and then
\beq
\label{piipp}
\pi^{(+)}_i(k) \equiv \frac{i\pi^{(+)}_{12}(k)}{2n_B(x)}\,,
\eeq
from which we obtain the damping rate of the $(+)$ mode by
means of \Eq{defGammamode}.  Moreover, from the result for $\pi^{(+)}_{i}$,
the corresponding contribution to the imaginary part of the
$V_L$ self-energy is
\beq
\pi^{(L)}_i(k) = |U_{2+}|^2 \pi^{(+)}_i(k)\,,
\eeq
and therefore, remembering \Eq{Imepsilonell}, we obtain for the
imaginary part of the \emph{longitudinal} dielectric constant
\beq
\label{imepsilon}
\Im \epsilon_{\ell}(\omega,\kappa) = 
-\frac{|U_{2+}|^2 \pi^{(+)}_i(k)}{k^2}\,.
\eeq

\section{Imaginary part of the self-energy - Calculation of $\pi^{(+)}_i$}
\label{sec-damping}

\subsection{Expression for $\pi^{(+)}_{i}$}

From \Fig{fig2},
\beq
\label{pi12pp}
-i(\pi^{(+)}_{12}(k)) = (iA)(-iA^\ast)\int\frac{d^4p}{(2\pi)^4}
i\Delta^{(-)}_{21}(p)i\Delta^{(-)}_{12}(k + p)\,,
\eeq
where $\Delta^{(-)}_{21}(q)$ and $\Delta^{(-)}_{12}(q)$ are given
in \Eq{tftpropagators}. Using the following identities
\beqa
\eta_B(q) + \theta(q) & = & e^{x_q} n_B(x_q)\epsilon(q)\,,\nonumber\\
\eta_B(q) + \theta(-q) & = & n_B(x_q)\epsilon(q)\,,
\eeqa
where
\beq
\epsilon(q) = \theta(q) - \theta(-q)\,,
\eeq
the propagators can be written in the form
\beqa
\label{tftpropagators2}
\Delta^{(s)}_{21}(q) & = & -2\pi i\delta(q^2 - \hat\sigma_{s}(q))
e^{x_q} n_B(x_q)\epsilon(q)\,,\nonumber\\
\Delta^{(s)}_{12}(q) & = & -2\pi i\delta(q^2 - \hat\sigma_{s}(q))
n_B(x_q)\epsilon(q)\,.
\eeqa
Then substituting \Eq{tftpropagators2} in \Eq{pi12pp}, after
some straightforward algebra [see \Appendix{app:Gamma+--}] we arrive at
\beqa
\label{piplusim}
\pi^{(+)}_i(k) & = & -\frac{1}{2}|A|^2
\int\frac{d^3P}{(2\pi)^3 2\omega^{(-)}_{r}(P)}
\frac{d^3Q}{(2\pi)^3 2\omega^{(-)}_{r}(Q)}(2\pi)^4\times\nonumber\\
&&\mbox{}
\Big\{\delta^{(4)}(k + p - q)(n_{-}(p) - n_{-}(q))\nonumber\\
&&\mbox{} + \delta^{(4)}(k + q - p)(-n_{-}(p) + n_{-}(q))\nonumber\\
&&\mbox{} - \delta^{(4)}(k + p + q)(n_{-}(p) + n_{-}(q) + 1)\nonumber\\
&&\mbox{} - \delta^{(4)}(k - p - q)(-n_{-}(p) - n_{-}(q) - 1)\Big\}\,.
\eeqa
In this formula,
\beqa
p^\mu & = & (\omega^{(-)}_r(P),\vec P)\,,\nonumber\\
q^\mu & = & (\omega^{(-)}_r(Q),\vec Q)\,,\nonumber\\
k^\mu & = & (\omega,\vec\kappa)\,.
\eeqa
\beq
\label{nspq}
n_{s}(p) \equiv \frac{1}{e^{\beta\omega^{(s)}_r(P)} - 1}\,,
\eeq
and similarly for $n_s(q)$ with $P\rightarrow Q$.
The damping of the propagating $(+)$ mode is then obtained by
\Eq{defGammamode}.

This expression for $\pi^{(+)}_i$ in \Eq{piplusim} can be interpreted
in terms of the transition probablities for varios processes,
\beqa
\phi_{+}(k) + \phi_{-}(p) & \longleftrightarrow & \phi_{-}(q) \,,\nonumber\\
\phi_{+}(k) + \phi_{-}(q) & \longleftrightarrow & \phi_{-}(p) \,,\nonumber\\
\phi_{+}(k) + \phi_{-}(q) + \phi_{-}(p) & \longleftrightarrow & 0\,,\nonumber\\
\phi_{+}(k) & \longleftrightarrow & \phi_{-}(q) + \phi_{-}(p)\,.
\eeqa
Consider, for example, the first term.  Using
\beq
n_{-}(p) - n_{-}(q) =
n_{-}(p)(1 + n_{-}(q)) - n_{-}(q)( 1 + n_{-}(p))\,,
\eeq
and remembering that the delta function fixes $k + p = q$,
the first term corresponds to the processes
\beq
\phi_{+}(k) + \phi_{-}(p) \longleftrightarrow \phi_{-}(q)\,.
\eeq
Some processes are forbiden by energy conservation while others
supppresed by our specific assumptions about the background conditions.

The dominant process in the case we consider is the decay process
$\phi_{+}(k) \rightarrow \phi_{-}(q) + \phi_{-}(p)$. For completeness
we also include its inverse, although it will be suppressed under the
background conditions that we assume. The corresponding contribution
to $\pi^{(+)}_i$ is 
\beq
\label{piplusim2}
\pi^{(+)}_i(k) = -\omega
\left[
\Gamma(\phi_{+}\rightarrow\phi_{-}\phi_{-}) -
\Gamma(\phi_{-}\phi_{-}\rightarrow \phi_{+})
\right]\,,
\eeq
where
\beqa
\label{Gammaforwback}
\Gamma(\phi_{+}\rightarrow\phi_{-}\phi_{-}) & = &
\frac{1}{2\omega}|A|^2
\int\frac{d^3P}{(2\pi)^3 2\omega^{(-)}_{r}(P)}
\frac{d^3Q}{(2\pi)^3 2\omega^{(-)}_{r}(Q)}(2\pi)^4
\delta^{(4)}(k - p - q)(1 + n_{-}(p))(1 + n_{-}(q))\,,\nonumber\\
\Gamma(\phi_{-}\phi_{-}\rightarrow\phi_{+}) & = &
\frac{1}{2\omega}|A|^2
\int\frac{d^3P}{(2\pi)^3 2\omega^{(-)}_{r}(P)}
\frac{d^3Q}{(2\pi)^3 2\omega^{(-)}_{r}(Q)}(2\pi)^4
\delta^{(4)}(k - p - q)n_{-}(p)n_{-}(q)\,.
\eeqa
In the limit that the $n_{-}$ distribution is negligible, the
dominant contribution to $\pi^{(+)}_i$ is due to the decay rate
\beq
\label{Gamma0neglegiblen}
\Gamma^{(0)}(\phi_{+}\rightarrow\phi_{-}\phi_{-}) =
\frac{1}{2\omega}|A|^2
\int\frac{d^3P}{(2\pi)^3 2\omega^{(-)}_{r}(P)}
\frac{d^3Q}{(2\pi)^3 2\omega^{(-)}_{r}(Q)}(2\pi)^4
\delta^{(4)}(k - p - q)\,.
\eeq
In any case, once we have computed $\pi^{(+)}_i$,
we obtain the damping of the $(+)$ mode from \Eq{defGammamode}.

\subsection{Long wavelength limit}

Here we will consider the calculation of $\pi^{(+)}_i$ in the
long wavelength limit ($\kappa\rightarrow 0)$, which is a useful
case for many applications.

\subsubsection{Evaluation of the integral}

We must evaluate the integral in \Eq{Gamma0neglegiblen} for $\kappa = 0$,
that is
\beq
I = \int\frac{d^3P}{(2\pi)^3 2\omega^{(-)}_{r}(P)}
\frac{d^3Q}{(2\pi)^3 2\omega^{(-)}_{r}(Q)}(2\pi)^4
\delta^{(4)}(k - p - q)\,,
\eeq
with
\beq
k^\mu = (\omega, \vec 0)\,,
\eeq
but using the exact dispersion relation for $\omega^{(-)}_r$. In this case we
can use the results of \Appendix{app:Ilongwavelength}, with the notation
adapted to the present case. As indicated at the end of
\Appendix{app:Ilongwavelength}, the results simplify in the present case
because $m_1 = m_2$. Thus, the result is
\beq
\label{I}
I = \sum_{P^\ast}\left(\frac{1}{2\pi}\right)^2
\frac{2\pi P^{\ast 2}}{\omega^2}
\frac{1}{|G_0|}\,,
\eeq
where $P^\ast$ is the solution of
\beq
\label{Pomegaeq}
\omega^{(-)}_{r}(P^\ast) = \frac{1}{2}\omega\,,
\eeq
and
\beq
\label{G0}
G_0 \equiv 
\left.\left(
\frac{\partial\omega^{(-)}_{r}}{\partial P}
\right)\right|_{P = P^\ast}\,.
\eeq
The $\sum_{P^\ast}$ symbol in \Eq{I} is meant to indicate a sum
over each valid solution of \Eq{Pomegaeq}. The identification of the valid
solutions is carried out below.

\subsubsection{Solution for $P^\ast$}

The full dispersion relation can be written in the form
\beq
\label{drm}
\omega^{(-)}_r = \sqrt{P^2 + M^2 - \sqrt{\Delta^2 + 4\mu^2 P^2}}\,,
\eeq
where we are defining
\beqa
\label{MDeltadef}
M^2 & \equiv \frac{1}{2}(m^2_\rho + m^2_V)\,,\nonumber\\
\Delta & \equiv \frac{1}{2}(m^2_\rho - m^2_V)\,.
\eeqa
We will solve \Eq{drm} for $P(\omega^{(-)}_r)$ and then
\beq
\label{Pastomega}
P^\ast(\omega) = P(\omega/2)\,.
\eeq

\begin{figure}
\begin{center}
\epsfig{file=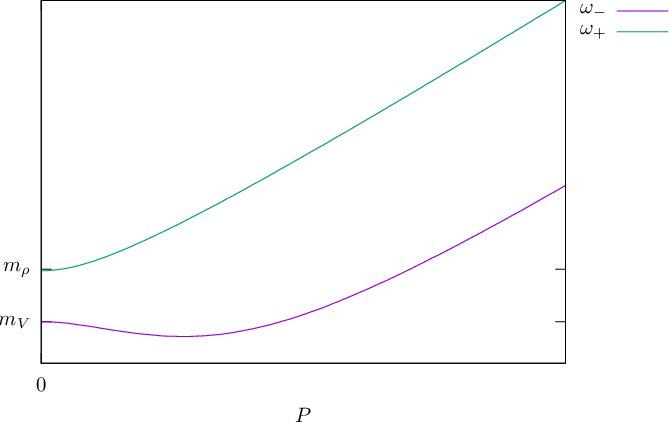,bbllx=163,bblly=293,bburx=484,bbury=496}
\end{center}
\caption[] {
  Sample sketches of the dispersions $\omega_{\pm}$ that illustrate
  the features discussed in the text and used in arriving at
  \Eq{Pastomegafinal}.
  \label{fig3}
}
\end{figure}

Solving \Eq{drm} for $P$ is tedious,
but straighforward since it is a quadratic equation.
The equation can be written in the form
\beq
P^4 - 2BP^2 - C = 0\,,
\eeq
where
\beqa
\label{BC}
B & = & 2\mu^2 - \left(M^2 - \omega^{(-)\,2}_r\right) \,,\nonumber\\
C & = & \Delta^2 - \left(M^2 - \omega^{(-)\,2}_r\right)^2\nonumber\\
& = & \left(\omega^{(-)\,2}_r - m^2_V\right)
\left(m^2_\rho - \omega^{(-)\,2}_r\right)\,.
\eeqa
Therefore the solutions are
\beq
\label{Pomegarm}
P^2 = B \pm \sqrt{D}\,,
\eeq
with
\beqa
\label{D}
D & = & B^2 + C\nonumber\\
& = & \Delta^2 + 4\mu^2\left[\mu^2 - M^2 + \omega^{(-)\,2}_r\right]\,.
\eeqa
For some values of $\omega^{(-)}_r$ both solutions in \Eq{Pomegarm}
are valid, but for others only $B + \sqrt{D}$ is valid.
As an aid to determine the ranges of $\omega^{(-)}_r$, we show a sketch
of the dispersion relations $\omega^{(\pm)}_r$ in \Fig{fig3}.

As the figure shows, the dispersion relation $\omega^{(-)}_r$ has a minimum
value, which we denote by $\omega^{(-)}_{r0}$. In that point the two
solutions for $P$ coincide, that is $D = 0$. Thus from \Eq{D},
\beq
\omega^{(-)\,2}_{r0} = M^2 - \mu^2 - \frac{\Delta^2}{4\mu^2}\,.
\eeq
For $\omega^{(-)}_{r0}< \omega^{(-)}_r < m_V$, it follows from \Eq{BC}
that $C < 0$. This implies that $\sqrt{D} < B$ and therefore both solutions
in \Eq{Pomegarm} are valid, as can be deduced also by a glance at
\Fig{fig3} in that region. In the region
$m_V < \omega^{(-)}_r < m_\rho$, we see from \Eq{BC} that $C > 0$,
and thefore only the solution $B + \sqrt{D}$ is valid.
Finally, for $\omega^{(-)}_r > m_\rho$,
it follows from \Eq{BC} that $C < 0$ and therefore both solutions
exist. However, as seen from \Fig{fig3}, the solution
with the smaller (larger) value of $P$ corresponds the $\omega^{(+)}_r$
($\omega^{(-)}_r$) dispersion relation (which is due to the fact
that $\omega^{(+)}_r$ increases faster than $\omega^{(-)}_r$
as $P$ increases.) Thus in that range the valid solution is
$B + \sqrt{D}$.

In summary,
\beqa
\label{Pastomegafinal}
P^{\ast\,2}(\omega) = \left\{
\begin{array}{ll}
B_\omega \pm \sqrt{D_\omega} & \quad
(\omega^{(-)}_{r0} < \omega/2 < m_V)\\[12pt]
B_\omega + \sqrt{D_\omega} & \quad (\omega/2 > m_V)
\end{array}\right.
\eeqa
where
\beqa
\label{BDomega}
B_\omega & = & 2\mu^2 - \left(M^2 - \frac{1}{4}\omega^2\right)\,,\nonumber\\
D_\omega & = &
\Delta^2 + 4\mu^2\left[\mu^2 - M^2 + \frac{1}{4}\omega^2\right]\,.
\eeqa

\subsubsection{Expression for
$\Gamma^{(0)}(\phi_{+}\rightarrow\phi_{-}\phi_{-})$}

From \Eq{drm}, it follows that
\beq
\left.\left(
\frac{\partial\omega^{(-)}_{r}}{\partial P}
\right)\right|_{P = P^\ast}
= \frac{2P^\ast}{\omega}
\left[1 - \frac{2\mu^2}{\sqrt{\Delta^2 + 4\mu^2 P^{\ast\,2}}}\right]\,,
\eeq
and therefore from \Eq{I}
\beq
\label{Ifinal}
I = \sum_{P^\ast}\frac{1}{4\pi}\left(\frac{P^\ast}{\omega}\right)
\left[1 - \frac{2\mu^2}{\sqrt{\Delta^2 + 4\mu^2 P^{\ast\,2}}}\right]^{-1}
\eeq

We now evaluate the parameters $U_{ab}$ and $A$, evaluated
at $\kappa = 0$. 
In the limit $\kappa \rightarrow 0$, the mixing parameters
defined in \Eq{Uai} become (remembering \Eq{Vrhohier}),
\beqa
\label{Uaikappa0}
U_{1+}(\kappa\rightarrow 0) & = & U_{2-}(\kappa\rightarrow 0) = 1\,,\nonumber\\
U_{1-}(\kappa\rightarrow 0) & = & \frac{2\mu\kappa}{m^2_\rho - m^2_V}
\,,\nonumber\\
U_{2+}(\kappa\rightarrow 0) & = &
-\frac{m_V}{m_\rho} U_{1-}(\kappa\rightarrow 0)\,,
\eeqa
which from \Eq{A} in turn give
\beq
A(\kappa\rightarrow 0) = A_1 + A_2\,,
\eeq
where
\beqa
A_1 & = & -q m_V[1 - 8rx^2]\,,\nonumber\\
A_2 & = & 8q\mu\left(\frac{\kappa}{\omega}\right)x[1 - 2rx^2]\,,
\eeqa
with
\beqa
r & \equiv & \frac{m_V}{m_\rho}\,,\nonumber\\
x & \equiv & \frac{\mu\kappa}{m^2_\rho - m^2_V}\,.
\eeqa
Thus, setting strictly $\kappa = 0$,
\beq
\label{A0}
A(0) = -q m_V\,.
\eeq
From \Eq{Gamma0neglegiblen}, we then finally obtain
\beq
\label{Gamma0}
\left.
\Gamma^{(0)}(\phi_{+}\rightarrow\phi_{-}\phi_{-})
\right|_{\kappa\rightarrow 0} =
\sum_{P^\ast}\frac{q^2 P^\ast}{8\pi}\frac{m^2_V}{\omega^2}
\left[1 -
\frac{2\mu^2}{\sqrt{\Delta^2 + 4\mu^2 P^{\ast\,2}}}
\right]^{-1}\,,
\eeq
which gives the leading term in \Eq{piplusim2} in the long wavelength
($\kappa\rightarrow 0$) limit.
The corrections due to the distribution functions
can be readily included using \Eq{Gammaforwback}, thus
\beq
\label{piplusimlw}
\pi^{(+)}_i(\omega,\kappa\rightarrow 0) = -\omega
\left.
\Gamma^{(0)}(\phi_{+}\rightarrow\phi_{-}\phi_{-})
\right|_{\kappa\rightarrow 0}
(n_{-}(p) + n_{-}(q) + 1)\,,
\eeq
remembering that in this limit
\beq
n_{-}(p) = n_{-}(q) = \frac{1}{e^{\frac{1}{2}\beta\omega} - 1}\,.
\eeq

\subsection{Damping in the long wavelenght limit}

The damping of the $(+)$ mode is given by \Eq{defGammamode}.
Explicitly, in this limit
\beq
\label{Gamma+final}
\Gamma_{+} = \frac{q^2 P^\ast_0}{8\pi}\frac{m^2_V}{m^2_\rho}
\left[1 - \frac{2\mu^2}{\sqrt{\Delta^2 + 4\mu^2 P^{\ast\,2}_0}}\right]^{-1}
\left[\frac{2}{e^{\frac{1}{2}\beta m_\rho} - 1} + 1\right]\,,
\eeq
where we have used the fact that in this limit
$\omega^{(+)}_r(\kappa\rightarrow 0) = m_\rho$, and
we have defined
\beq
P^{\ast}_0 \equiv P^\ast(m_\rho)\,.
\eeq
From \Eqs{Pastomegafinal}{BDomega}, using \Eq{MDeltadef},
\beq
\label{Pastdamping}
P^{\ast\,2}_0 = 2\mu^2 - \frac{1}{4}(m^2_\rho + 2m^2_V) +
\sqrt{4\mu^4 + \frac{1}{4}(m^2_\rho - m^2_V)^2 - \mu^2(m^2_\rho + 2m^2_V)}\,,
\eeq
provided that
\beq
\label{mrhomVcondition}
m_\rho > 2m_V\,.
\eeq

\subsection{Longitudinal dielectric constant}

The corresponding contribution to the longitudinal dielectric constant
is given by \Eq{imepsilon}. Therefore, using
\Eqs{Gamma0}{piplusimlw}, together with \Eq{Uaikappa0},
\beq
\label{epsilonellfinal}
\Im\epsilon_{\ell} = \sum_{P^\ast}
\frac{q^2}{8\pi}
\frac{P^\ast(\omega)\kappa^2}{\omega^3}
\frac{m^2_V}{m^2_\rho}\left(
\frac{2\mu m_V}{m^2_\rho - m^2_V}
\right)^2
\left[
1 - \frac{2\mu^2}{\sqrt{\Delta^2 + 4\mu^2 P^{\ast\,2}(\omega)}}
\right]^{-1}
\left[\frac{2}{e^{\frac{1}{2}\beta\omega} - 1} + 1\right]\,,
\eeq
which complements \Eq{Reepsilonell}.

\section{Discussion}
\label{sec:discussion}

For guidance and illustrative purposes, let us consider the
large $m_\rho$ limit. There are three indendent parameters,
\beq
\mu, m_\rho, m_V\,.
\eeq
Remember that,
\beqa
m^2_\rho & = & 4\mu^2 + 2(\mu^2 - m^2)\nonumber\\
m^2_V & = & \frac{q^2}{\lambda}(\mu^2 - m^2)\,.
\eeqa
So we can think also in terms of
\beq
\mu, m, \frac{q^2}{\lambda}\,.
\eeq
For the purposes of this discussion we will take
\beqa
\frac{q^2}{\lambda} & \sim & 1\nonumber\\
\mu & \sim & m\,,
\eeqa
so that, in particular,
\beq
\label{mrhomuapprox}
m_\rho \sim 2\mu \sim 2m \gg m_V\,.
\eeq

\subsection{Damping}

From \Eq{Pastdamping}, neglecting $m_V$,
\beq
P^{\ast\,2}_0 \simeq \frac{3}{4}m^2_\rho\,,
\eeq
and therefore we will put
\beq
P^{\ast}_0 \sim m_\rho\,.
\eeq
Then, under the conditions of \Eq{mrhomuapprox},
in the $m_V\rightarrow 0$ limit \Eq{Gamma+final} gives
\beq
\Gamma_{+} \simeq \frac{q^2}{4\pi}\frac{m^2_V}{m_\rho}\,,
\eeq
neglecting the factor involving the distribution function.

\subsection{Imaginary part of the longitudinal dielectric constant}

We consider \Eq{epsilonellfinal} in the same limit discused above.
Since we are considering $m_V \rightarrow 0$, the appropriate
formula to use for $P^{\ast\,2}(\omega)$ is the second one in
\Eq{Pastomegafinal} (i.e., $\omega/2 > m_V$),
which in the $m_V \rightarrow 0$ limit reduces to
\beq
\label{PastsmallmV}
P^{\ast\,2}(\omega) = \frac{1}{4}\omega^2 + \frac{1}{2}m_\rho\omega\,.
\eeq
In the limit stated in \Eq{mrhomuapprox}, the formula in \Eq{epsilonellfinal}
then gives
\beq
\label{epsilonellsmallV}
\Im\epsilon_{\ell} = \frac{q^2}{8\pi}
\frac{P^\ast(\omega)\kappa^2}{\omega^3}
\left(\frac{m_V}{m_\rho}\right)^4
\left[
1 - \frac{1}{\sqrt{1 + \frac{4P^{\ast\,2}(\omega)}{m^2_\rho}}}
\right]^{-1}\,,
\eeq
neglecting the correction due to the distribution functions.
Further insight can be obtained by considering the cases $\omega \gg m_\rho$
or $\omega \ll m_\rho$. From \Eq{PastsmallmV},
\beq
\frac{P^{\ast\,2}(\omega)}{m^2_\rho} =
\left\{\begin{array}{ll}
\frac{\omega^2}{4m^2_\rho} & (\omega \gg m_\rho)\,,\\[12pt]
\frac{\omega}{2m_\rho} & (\omega \ll m_\rho)\,,
\end{array}\right.
\eeq
and correspondingly
\beq
\left[1 - \frac{1}{\sqrt{1 + \frac{4P^{\ast\,2}(\omega)}{m^2_\rho}}}
\right]^{-1} =
\left\{\begin{array}{ll}
1 + O\left(\frac{m_\rho}{\omega}\right) & (\omega \gg m_\rho)\,,\\[12pt]
\frac{m_\rho}{\omega} & (\omega \ll m_\rho)\,.
\end{array}\right.
\eeq
Therefore,
\beq
\Im\epsilon_{\ell} = \frac{q^2}{8\pi}
\left(\frac{m_V}{m_\rho}\right)^4
\left(\frac{\kappa}{m_\rho}\right)^2
\times
\left\{\begin{array}{ll}
\frac{m^2_\rho}{2\omega^2} & (\omega \gg m_\rho)\,,\\[12pt]
\frac{1}{\sqrt{2}}\left(\frac{m_\rho}{\omega}\right)^{7/2} &
(\omega \ll m_\rho)\,.
\end{array}\right.
\eeq

\subsection{Real part of the longitudinal dielectric constant}

In the same limit we have considered above, more specifically
\beq
\label{mrhomuapprox2}
m_\rho \sim 2\mu \sim 2m \gg m_V,\omega,\kappa\,,
\eeq
it follows from \Eq{Reepsilonell} that
\beq
\label{ReepsilonellLmrho}
\Re\epsilon_\ell = 1 - \frac{m^2_V}{\omega^2 - \frac{\kappa^4}{m^2_\rho}}\,.
\eeq
The details leading to \Eq{ReepsilonellLmrho} are given in
\Appendix{app:largemrho}. Here we note that the dispersion relation
that follows from
\beq
\Re\epsilon_\ell = 0\,,
\eeq
(neglecting the damping term) is simply
\beq
\label{omegamLmrho}
\omega^2_V = m^2_V + \frac{\kappa^4}{m^2_\rho}\,.
\eeq
This is the same result that is obtained from \Eq{omegapm}
for $\omega^{(-)}_r$ in the same limit\cite{ns:gbec}. Furthermore,
it is reasuring that, putting $m_\rho \sim 2m$, as indicated in
\Eq{mrhomuapprox2}, \Eqs{ReepsilonellLmrho}{omegamLmrho} reproduce
the results obtained for the dielectric constant and dispersion relation
in non-relativistic models of the BE condensation of charged scalars\cite{2prb}.

\begin{figure}
\begin{center}
\epsfig{file=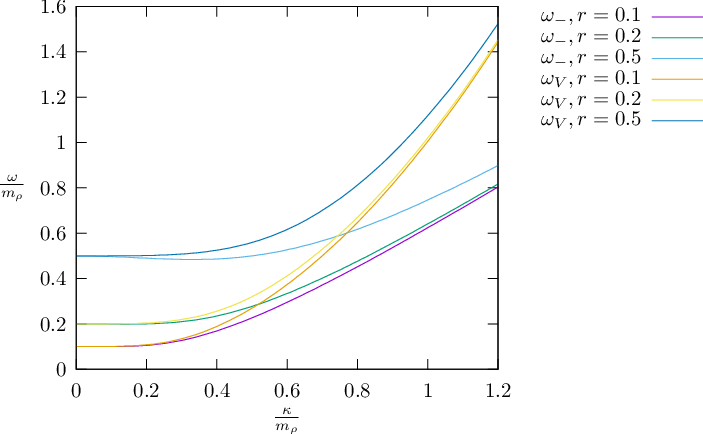,bbllx=146,bblly=290,bburx=484,bbury=499}
\end{center}
\caption[] {
   Plots of the dispersion relations $\omega^{(-)}_r$ and $\omega_V$, given in
   \Eqs{omegapm}{omegamLmrho}, respectively. For the purpose of the plots we
   have set $2\mu/m_\rho = 1$ [see \Eq{mrhomuapprox2}]. 
   The dispersion relations are ploted for the values of the parameter
   $r \equiv \frac{m_V}{m_\rho}$, as indicated.
  \label{fig4}
}
\end{figure}
However it must be kept in mind that the results given in
\Eqs{ReepsilonellLmrho}{omegamLmrho} are limited by the fact that
they are valid and hold in the limit of large $m_\rho$.
In this respect, \Eq{Reepsilonell} gives the appropriate formula to
use for the longitudinal dielectric constant when this limiting case
is not apropriate, and correspondingly \Eq{omegapm} for the dispersion relation
$\omega^{(-)}_r$.

For illustrative purposes and guidance, \Fig{fig4} shows the plots of the
dispersion relations $\omega^{(-)}_r$ and $\omega_V$, given in
\Eqs{omegapm}{omegamLmrho}, respectively. For the purpose of the plots we
have set $2\mu/m_\rho = 1$ [see \Eq{mrhomuapprox2}]. The dispersion relations
are ploted for various values of the parameter
$r \equiv \frac{m_V}{m_\rho}$ indicated in the figure.

\section{Conclusions and Outlook}
\label{sec:conclusions}

The subject of the present work is the imaginary part of the dispersion
relations of the propagating modes in a model of a charged scalar
Bose-Einstein (BE) condensate.
In a previous work\cite{ns:gbec} we considered the real part of the dispersion
relations. In that model, two modes correspond to the transverse photon
polarizations, while the other two modes are combinations
of the longitudinal photon and the massive scalar field, which we denote
as the $(\pm)$ modes. While the dispersion relations of the transverse modes
have the usual form for transverse photons in a plasma, the dispersion
relations of the ($\pm$) modes reveal some unique features. In our
previous work cited above, we determined and analyzed the real part of the
dispersion relations of the $(\pm)$ modes, as well as the real part of
the contribution to the longitudinal component of polarization tensor
and the corresponding contribution to the dielectric constant. The focus
in the present work is the calculation, in the same model, of
the imaginary part of the dispersion relations of the ($\pm$) modes,
as well as the corresponding contributions to the imaginary part
of the longitudinal component of polarization tensor
and the dielectric constant.

The main result of the present work is an explicit formula for the
imaginary part of the dispersion relations,
as well as the corresponding contributions to the imaginary part
of the longitudinal component of polarization tensor
and the dielectric constant, obtained by the one-loop
calculations of the self-energy of the propagating ($\pm$) modes
in the context of TFT. To illustrate some features of the results
we considered in some detail the long wavelength limit, including
specifically the high and low frequency limits, and obtained explicit
simple explicit expressions for the damping rate as well as the
background contribution ot the longitudinal polarization
tensor and dielectric constant in those cases. This work and the results
obtained pave the way for further exploration of the
properties of the model, which can be useful for
aplications related to the optical properties of this kind
of system in various physical contexts, such as those mentioned
in the references already cited, and experiments related
to dark-matter searches using materials with special optical
properties\cite{dmsearches}.

\section*{Appendix}
\appendix

\section{Details of \Eq{piplusim}}
\label{app:Gamma+--}

After substituting \Eq{tftpropagators2} in \Eq{pi12pp} we use
\beq
e^{x_p}n_B(x_p)n_B(x_{p + k}) =
n_B(x_k)\left(n_B(x_p) - n_B(x_{p + k})\right)\,.
\eeq
Then from \Eq{piipp}, we obtain
\beqa
\pi^{(+)}_i(k) & = & -2\pi^2 |A|^2\int\frac{d^4p}{(2\pi)^4}
\delta(p^2 - \hat\sigma_{-}(p))\delta((p + k)^2 - \hat\sigma_{-}(p + k))
\nonumber\\
&&\mbox{}\times
\left(n_B(x_p) - n_B(x_{p + k})\right)\epsilon(p)\epsilon(p + k)\,.
\eeqa
We now put
\beq
p + k = q\,,
\eeq
and insert
\beq
(2\pi)^4\int\frac{d^4q}{(2\pi)^4}\delta^{(4)}(q - p - k)\,.
\eeq
Thus,
\beqa
\pi^{(+)}_i(k) & = & -\frac{|A|^2}{2}
\int\frac{d^4p}{(2\pi)^3}\int\frac{d^4q}{(2\pi)^3} 
(2\pi)^4\delta^{(4)}(q - p - k)
\delta(p^2 - \hat\sigma_{-}(p))\delta(q^2 - \hat\sigma_{-}(q))
\nonumber\\
&&\mbox{}\times
\left(n_B(x_p) - n_B(x_q)\right)\epsilon(p)\epsilon(q)\,.
\eeqa
Using
\beq
\epsilon(p)\epsilon(q) =
\theta(p)\theta(q) + \theta(-p)\theta(-q)
- \theta(p)\theta(-q) - \theta(-p)\theta(q)\,,
\eeq
and
\beq
n_B(x) = -1 - n_B(-x)\,,
\eeq
we have
\beqa
\left(n_B(x_p) - n_B(x_q)\right)\epsilon(p)\epsilon(q) & = &
\theta(p)\theta(q)\left(n_B(x_p) - n_B(x_q)\right)\nonumber\\
&&\mbox{} + \theta(-p)\theta(-q)\left(-n_B(-x_p) + n_B(-x_q)\right)\nonumber\\
&&\mbox{} - \theta(p)\theta(-q)\left(n_B(x_p) + 1 + n_B(-x_q)\right)\nonumber\\
&&\mbox{} - \theta(-p)\theta(q)\left(-1 - n_B(-x_p) - n_B(x_q)\right)\,.
\eeqa
We will then use
\beqa
\int\frac{d^4p}{(2\pi)^3}\delta(p^2 - \hat\sigma_{s}(p))\theta(p)n_B(x_p) & = &
\int\frac{d^3P}{(2\pi)^3 2\omega^{(s)}_r(P)} n_s(p)\,,\nonumber\\
\int\frac{d^4p}{(2\pi)^3}\delta(p^2 - \hat\sigma_{s}(p))\theta(-p)n_B(-x_p)
& = & \int\frac{d^3P}{(2\pi)^3 2\omega^{(s)}_r(P)} n_s(p)\,,
\eeqa
where
\beq
n_{s}(p) \equiv \frac{1}{e^{\beta\omega^{(s)}_r(P)} - 1}\,,
\eeq
and with the understanding that, after the integration over $p^0$,
\beq
p^\mu = (\omega^{(s)}_r(P),\vec P)\,.
\eeq
Similarly with the integrals over $q$, with
$q^\mu = (\omega^{(s)}_r(Q),\vec Q)$.
The functions $\omega^{(s)}_r(P)$ and $\omega^{(s)}_r(Q)$ are the
dispersion relation defined in \Eq{omegapm}.

In this way we arrive at
\beqa
\pi^{(+)}_i(k) & = & -\frac{1}{2}|A|^2
\int\frac{d^3P}{(2\pi)^3 2\omega^{(-)}_{r}(P)}
\frac{d^3Q}{(2\pi)^3 2\omega^{(-)}_{r}(Q)}(2\pi)^4\times\nonumber\\
&&\mbox{}
\Big\{\delta^{(4)}(k + p - q)(n_{-}(p) - n_{-}(q))\nonumber\\
&&\mbox{} + \delta^{(4)}(k + q - p)(-n_{-}(p) + n_{-}(q))\nonumber\\
&&\mbox{} - \delta^{(4)}(k + p + q)(n_{-}(p) + n_{-}(q) + 1)\nonumber\\
&&\mbox{} - \delta^{(4)}(k - p - q)(-n_{-}(p) - n_{-}(q) - 1)\Big\}\,,
\eeqa
with
\beqa
p^\mu & = & (\omega^{(-)}_r(P),\vec P)\,,\nonumber\\
q^\mu & = & (\omega^{(-)}_r(Q),\vec Q)\,,\nonumber\\
k^\mu & = & (\omega,\vec\kappa)\,,
\eeqa
which is \Eq{piplusim}.

\section{Phase space integral for $\phi_{+} \rightarrow \phi_{-}\phi_{-}$
  in the long wavelength limit}
\label{app:Ilongwavelength}
  
We consider the phase space integral of the type we have to do
for the $\phi_{+} \rightarrow \phi_{-}\phi_{-}$ decay, that is
\beq
\label{Ipmm}
I \equiv
\int\frac{d^3 P_1}{2E_1} \int\frac{d^3 P_2}{2E_2}
\delta^{(4)}(k - p_1 - p_2)\,,
\eeq
but with the dispersion relations of the form
\beq
\label{E1P1}
E_i(P) = \sqrt{P^2 + m^2_i(P)}\,.
\eeq
In this section we consider the long wavelength limit, which
corresponds to setting $\kappa\rightarrow 0$,
\beq
\label{kkappazero}
k^\mu = (\omega, \vec 0)\,.
\eeq

We start from
\beq
I =\int\frac{d^3P_1}{2E_1}\delta((k - p_1)^2 - m^2_2)\theta(\omega - E_1)\,,
\eeq
where now
\beq
(k - p_1)^2 = \omega^2 + m^2_1 - 2\omega E_1\,,
\eeq
and
\beq
p_2 = k - p_1 \qquad \Rightarrow \qquad P_2 = P_1\,.
\eeq
Then putting
\beq
\frac{d^3P_1}{2E_1} \rightarrow 2\pi\frac{P^2_1}{E_1} dP_1\,,
\eeq
we obtain
\beqa
\label{dIkappazero}
dI & = & 2\pi dP_1 \frac{P^2_1}{E_1}
\delta[\omega^2 + m^2_1(P_1) - m^2_2(P_1) - 2\omega E_1]\theta(\omega - E_1)
\nonumber\\
& = & \frac{\pi}{\omega} dP_1 \frac{P^2_1}{E_1}
\delta[\omega^2 + m^2_1(P_1) - m^2_2(P_1) - 2\omega E_1]\theta(\omega - E_1)\,.
\eeqa
The delta function then fixes $P_1 = P^\ast_1$, where $P^\ast_1$ is such that
\beq
E_1(P^\ast_1) = \frac{\omega^2 + m^2_1(P^\ast_1) - m^2_2(P^\ast_1)}{2\omega}\,,
\eeq
with $E_1(P_1)$ given in \Eq{E1P1}. This is an implicit equation that must
be solved for $P^\ast_1$. The theta function $\theta(\omega - E_1)$ requires
that $\omega$ and the parameters that enter in the definition of the $m_i(P)$
be such that
\beq
E_1(P^\ast_1) < \omega
\eeq
otherwise the integral is zero (this is the analog of the condition
$m > m_1 + m_2$ in the standard case). Assuming this is satisfied, then
\beq
I = \frac{\pi}{\omega}\frac{P^{\ast 2}_1}{E^\ast_1}
\frac{1}{|G_0|}\,,
\eeq
where we have defined $E^\ast_1$,
\beq
E^\ast_1 \equiv E_1(P^\ast_1) =
\frac{\omega^2 + m^2_1(P^\ast_1) - m^2_2(P^\ast_1)}{2\omega}\,,
\eeq
and
\beq
G_0 \equiv 
\left.\left(\frac{\partial G}{\partial P_1}\right)\right|_{P_1 = P^\ast_1}\,,
\eeq
with
\beq
G = E_1(P_1) + \frac{m^2_1(P_1) - m^2_2(P_1)}{2\omega}\,.
\eeq

\subsection*{Special case: $m_1 = m_2$}

In this case the formulas simplify. The equation for $P^\ast_1$ is simply
\beq
E_1(P^\ast_1) = \frac{1}{2}\omega\,.
\eeq
Assuming that $\omega$ and the parameters that enter in the definition
of $m_1(P)$ are such that the solution exists, the condition
$E_1(P^\ast_1) < \omega$ it automatically satisfied.
$G$ reduces to $E_1(P)$. Therefore,
\beq
I = 2\pi\frac{P^{\ast 2}_1}{\omega^2}
\frac{1}{|G_0|}\,,
\eeq
with
\beq
G_0 =
\left.\left(\frac{\partial E_1}{\partial P_1}\right)\right|_{P_1 = P^\ast_1}\,.
\eeq

\section{Derivation of \Eq{ReepsilonellLmrho}}
\label{app:largemrho}

Rewriting the term in the square brackets in \Eq{ReepsilonellLmrho},
\beqa
\Re\epsilon_\ell & = & 1 - \frac{m^2_V(k^2 - m^2_\rho)}
{k^2(k^2 - m^2_\rho) - 4\mu^2\kappa^2}\nonumber\\
& = & 1 - \frac{m^2_V}{X}\,,
\eeqa
where
\beq
X \equiv
k^2 + \frac{4\mu^2\kappa^2}{m^2_\rho\left(1 - k^2/m^2_\rho\right)}\,.
\eeq
Expanding the denominator,
\beq
X = \omega^2\left(1 + \frac{\eta\kappa^2}{m^2_\rho}\right) + \kappa^2(\eta - 1)
- \frac{\eta\kappa^4}{m^2_\rho}\,,
\eeq
where we have put
\beq
\eta = \frac{4\mu^2}{m^2_\rho}\,.
\eeq
In the limit stated in \Eq{mrhomuapprox}, $\eta \rightarrow 1$, and
\beq
X \rightarrow \omega^2 - \frac{\kappa^4}{m^2_\rho}\,,
\eeq
which leads to \Eq{ReepsilonellLmrho}.

%\bibliographystyle{ieeetr}
%\bibliography{main}

\end{document}